\documentclass[11pt,twoside]{article}

\usepackage{asp2006}
\usepackage{natbib}
\begin{document}
\title{Planetesimal Formation with Particle Feedback}   %%% Fill in title
\author{Andrew N. Youdin}   %%% Fill in author names
\affil{Canadian Institute for Theoretical Astrophysics,  University of Toronto, 60 St.\ George St., Toronto, ON M5S 3H8, Canada}    %%% Fill in author affiliations
\author{Anders Johansen}   %%% Fill in author names
\affil{Max-Planck-Institut f\"ur Astronomie, 69117 Heidelberg, Germany}    %%% Fill in author affiliations
\begin{abstract} %%% Abstract to run on from here.
Proposed mechanisms for the formation of km-sized solid planetesimals face long-standing difficulties.  Robust sticking mechanisms that would produce planetesimals by coagulation alone remain elusive.  The gravitational collapse of smaller solids into planetesimals is opposed by stirring from turbulent gas.   This proceeding describes recent works showing that ``particle feedback," the back-reaction of drag forces on the gas in protoplanetary disks, promotes particle clumping as seeds for gravitational collapse.  The idealized streaming instability demonstrates the basic ability of feedback to generate particle overdensities.  More detailed numerical simulations show that the particle overdensities produced in turbulent flows trigger gravitational collapse to planetesimals.  We discuss surprising aspects of this work, including the large (super-Ceres) mass of the collapsing bound cluster, and the finding that MHD turbulence aids gravitational collapse.
\end{abstract}

%%% MAIN BODY OF TEXT GOES HERE. CONSULT "INSTRUCTIONS FOR AUTHORS USING
%%% LATEX2E MARKUP", SECTIONS 2.3-2.6 FOR HELP WITH EQUATIONS, FIGURES,
%%% AND TABLES.

%\section{}   %%% Top level section head (remove "%" symbol)
%\subsection{}   %%% Second level section head (remove "%" symbol)
%\subsubsection{}   %%% Lowest level section head (remove "%" symbol)
%\section*{}    %%% Unnumbered top level section head (remove "%" symbol)
%\subsection*{}   %%% Unnumbered second level section head (remove "%" symbol)

\section{Introduction}
Coagulation is the dominant mechanism for the growth of dust grains via van der Waals forces  \citep{dt97} and for the growth of solid protoplanets by gravitational binding \citep{gls04}.  But sticking is difficult in the ~mm--km size range, as confirmed by an extensive body of experimental work \citep{wb06}.  Moreover, incremental growth of planetesimals leads to the rapid ($\sim10^2$ yr)  inspiral of particles near a meter in size.  

An alternative hypothesis \citep[hereafter GW]{saf69, gw73} proposes that km-sized planetesimals formed from the gravitational collapse of smaller solids.  This mechanism overcomes the sticking and radial drift obstacles in one fell swoop.  However stirring by turbulent gas opposes gravitational collapse.   
GW noted that their dense particle midplane would trigger a turbulent boundary layer via Kelvin-Helmholtz instabilities.    \citet{sw80} argued that this particle-driven turbulence (an example of drag force feedback) would stir up the particle midplane enough to prevent gravitational collapse.  This appeared to be a fatal flaw of the GI hypothesis, since particle settling was a self-limiting process.

\citet{sek98} and \citet{ys02} showed that particles could actually help their cause of becoming planetesimals.  When the surface density  of solids relative to gas is larger (by factors of few) than solar abundances, then vertical shear is no longer strong enough to overcome the anti-buoyancy of the dense midplane layer.  The ability of particles to stir themselves is limited.

However these works did not model the back reaction of drag forces on the gas in detail.  Indeed most analyses of midplane Kelvin-Helmholtz instabilities \citep[with the exception of][]{jhk06} reduce the particle and gas dynamics to a simplified set of equations which omit relative motion and prevent a detailed examination of the effects of drag forces.

Section \ref{sec:SI} describes the streaming instability \citep[hereafter YG]{yg05} which takes the opposite approach of neglecting stratification and vertical shear to consider two-way drag forces in a self-consistent, if incomplete, model.  This idealized system shows that particle feedback triggers spontaneous particle clumping.
The detailed 3D simulations of  \citet[hereafter JOMKHY]{nature07} include vertical stratification, self-gravity, MHD turbulence and multiple particle sizes, see \S\ref{sec:KS}  These models show that clumps of 15 -- 60 cm boulders, augmented by feedback effects, collapse gravitationally into bound clusters, which should continue to contract into planetesimals.

\section{The Streaming Instability}\label{sec:SI}
\subsection{Linear Growth}
The streaming instability arises in a simplified model of midplane layers in a protoplanetary disk (YG).  The model analyzes the local, axisymmetric Keplerian dynamics of gas and a single particle species in the absence of stratification.  The gas and solid components are coupled by a linear drag acceleration which satisfies Newton's third law by including the back-reaction of drag forces on the gas.  The only forcing is a constant radial acceleration, representing the global pressure gradient, which induces sub-Keplerian rotation of the gas.

Due to the neglect of stratification and particle settling, the model has a well-defined equilibrium state.  The sub-Keplerian headwind robs the particles of angular momentum and  causes an inward (outward) radial drift for solids (gas).  The coupled drift velocities are \citep[hereafter YJ]{nsh86, yj07},
\begin{eqnarray}
  u_x &=& \,\,\, \frac{2 \mu \tau_{\rm s}}
      {(1+\mu)^2+\tau_{\rm s}^2} \eta v_{\rm K}
      \, , \label{eq:NSHux} \\
  u_y &=& -\left[1+ \frac{\mu \tau_{\rm s}^2}
      {(1+\mu)^2+\tau_{\rm s}^2} \right]\frac{\eta v_{\rm K}}{1+\mu}
      \, , \label{eq:NSHuy} \\
  w_x &=& -\frac{2 \tau_{\rm s}}
      {(1+\mu)^2+\tau_{\rm s}^2} \eta v_{\rm K}
      \, , \label{eq:NSHwx} \\
  w_y &=& -\left[1- \frac{ \tau_{\rm s}^2}
      {(1+\mu)^2+\tau_{\rm s}^2} \right]\frac{\eta v_{\rm K}}{1+\mu}
      \, , \label{eq:NSHwy}
\end{eqnarray} 
$u_z = w_z = 0$, where $u_i$ ($w_i$) are the velocity components of gas (particles) relative to a Keplerian orbit, and $(x,y,z)$ are the locally Cartesian radial, azimuthal, and vertical coordinates.  The velocity scale, $\eta v_{\rm K} \sim 20$ --  50 m/s is set by the global radial pressure gradient (see YG or YJ).   The system is characterized by two dimensionless parameters: $\tau_{\rm s} \equiv \Omega t_{\rm stop}$, the aerodynamic stopping time $t_{\rm stop}$ normalized to the Keplerian orbital frequency, $\Omega$, is a proxy for particle size (bigger particles give larger values of $\tau_{\rm s}$); and $\mu = \rho_{{\rm p}}/\rho_{{\rm g}}$, the equilibrium  particle-to-gas density ratio.\footnote{In YJ and JY, $\epsilon$ was used in place of $\mu$ for consistency with previous numerical simulations.}

%physical description of instability
YG found that equilibrium drift solutions are robustly unstable, with linear growth for any values of $\tau_{\rm s}$ and $\mu$.
The instability produces particle overdensities as a mechanism to communicate the back-reaction of the drag force on the gas.  Gas density perturbations exist, but with much smaller amplitudes since motions are subsonic.

The robustness of the instability relies on the free energy contained in the relative drift of particles and gas.  Unlike the two stream instability in plasma physics, which is mediated by electric fields instead of dissipative drag forces, rotation is required to drive streaming instabilities.  Ultimately the instability is driven by the work done by the global pressure gradient on the outward drifting gas.  Thus the streaming instability differs from (e.g.)\ the magneto-rotational instability \citep{bh91}, which taps the energy released by the outward transport of angular momentum due to orbital (usually Keplerian) shear.  See YJ\S 5  for an analysis of disk energetics with particle-gas coupling.

%simple system gives complex behavior
Though the input physics to the streaming instability is rather simple, the resulting behavior is quite complex, partly because the motions contain many degrees of freedom.  Axisymmetric growth requires a ``2.5D" analysis, with all three velocity components included for each species plus particle density and gas pressure (for a total of 8 components). 

The peak growth rate increases with $\mu$ as the strength of feedback increases.  One might expect that growth rates would peak near $\tau_{\rm s} = 1$, where most drag related processes (e.g.\ drift speeds and vortex trapping) are maximized.  This is indeed the case in the ``normal" gas-dominated regime, $\mu < 1$.  However in the particle-dominated regime, $\mu > 1$ a curious behavior occurs.  The peak growth rate increases as $\tau_{\rm s} \rightarrow 0$.  This behavior is seen in Fig.\ 3 of YG and Fig.\ 2 of YJ.  The rapid growth for $\tau_{\rm s} \ll 1$ with $\mu > 1$ is only possible because it occurs on small length scales.

%Thus two practical obstacles exists for rapid streaming instabilities for small particles with $\tau_{\rm s} \ll 1$.  First, any anomalous viscosity, such as from turbulent diffusion (or from grid effects in a numerical simulation), could damp the small scale modes which give rapid growth.  Second, while $\mu > 1$ can simply be assumed as an initial state, reaching the particle-dominated regime is difficult in practice for small particles.  Small particles settle slowly and tend to clump less in MHD turbulence \citep{jkh06}, but could concentrate at the dissipation scale \citep{max87,cuz01}.

\subsubsection*{Numerical Confirmation}

The linear growth rates derived by YG were confirmed by numerical simulations in YJ.  An Eulerian hydrodynamics code (the Pencil Code) was used to model the gas, while Lagrangian ``super-particles" (because a realistic number of particles is impossible to include) represented the solid component.  Reproducing the linear growth rates by careful seeding of velocity and density eigenmodes was useful for calibration (and algorithm development) of the code.  This task also confirmed that the analytic treatment of particles as a pressureless fluid in YG was a valid analytical approximation.

\subsection{Non-linear Evolution}\label{sec:SInonlin}
The non-linear evolution of the streaming instability into particle-driven turbulence was explored in \citet[hereafter JY]{jy07}, using the same code as YJ.  The qualitative behavior was quite diverse, depending on the parameter values.  In simulations with $\tau_{\rm s} = 1$ (roughly 20 cm -- m sized boulders), the linear growth phase was followed by an upward cascade to large clumps which survived for tens of orbits where particles collected as in a traffic jam.  Smaller particles, with $\tau_{\rm s} = 0.1$ did not show a pronounced upward cascade and clumps were generated and destroyed in an orbital time.  A remarkable surprise (in the $\tau_{\rm s} = 0.1$ runs) was that the densest particle clumps drifted inwards faster than the particles in voids, a reversal of the density dependence for laminar flows.

The most interesting statistic (for the purposes of planetesimal formation) is the degree of particle clumping.  The large particle ($\tau_{\rm s} = 1$) runs showed the greatest clumping, with peak particle overdensities of several hundred.  The 
small particle ($\tau_{\rm s} = 0.1$) runs produced particle densities up to factors of tens, as long as the mass ratio $\mu \geq 1$.  Though the linear instability is very powerful in this regime, the shorter clump lifetime produces more modest (but still highly non-linear) overdensities.  Strongly non-linear clumping is not universal.  In the small particle, gas-dominated case ($\mu = 0.2$, $\tau_{\rm s} = 0.1$) then particle density fluctuations were order unity at most, consistent with the slow growth rates in this regime.

While peak densities are physically relevant (e.g.\ for gravitational collapse), they are always numerically underestimated, as higher resolution and larger simulation boxes will find rarer fluctuations.  JY also analyzed the transfer of angular momentum (which was inward!) and the diffusion of particles.

\section{The Kitchen Sink Simulations}\label{sec:KS}
While the streaming instability provides an idealized demonstration of the relevance of particle feedback, the simulations in JOMKHY included a much wider range of physical processes, hence the tongue-in-cheek label ``kitchen sink."

\subsection{Feedback and MHD turbulence}\label{sec:KSclump}
The first set of simulations in JOMKHY (see their Fig.\ 1) studied the effect of feedback on particle clumping in magneto-rotational turbulence.  Simulations were fully 3D with ideal MHD.  Vertical gravity was included to allow particle settling (but was not necessary, or included, for the gas since only the midplane region was considered).  Without feedback, particles (with uniform $\tau_{\rm s} = 1$) reached a peak density of 10 -- 30 times the gas density due to a balance between vertical settling, pressure trapping, and (on the other hand) turbulent diffusion.  This concentration factor is $> 10^3$ relative to uniform mixing with the gas (since the mass fraction of solids-to-gas was fixed at the solar abundance of 1\%).\footnote{Recall the comment in \S\ref{sec:SInonlin} that the peak particle densities are underestimates and are most meaningful for comparing simulations at the same resolution.} 

When the back-reaction of the drag force on the gas is included, peak particle densities were amplified by a further factor of 10 (to $> 100$ times the gas density).  The ability of particle feedback to enhance particle clumping is robust, and not just a consequence of the most idealized case of the streaming instability.  These simulations  with feedback included all viable sources of turbulence;  externally imposed (here magneto-rotational), and particle driven both by vertical shear and streaming instabilities; and still particle clumping is strong.

\subsection{Gravitational Collapse Simulation}
The particle clumping discussed above should readily trigger gravitational collapse.  Particle densities exceeding the gas density by $\ga 100$ also exceed the Roche density, since $\rho_{\rm Roche}/\rho_{\rm g} \simeq \Omega^2/(G \rho_{\rm g}) \sim 20$ (for double the minimum solar nebula as in the simulations).  To test this expectation, self-gravity was added to the model.  For greater realism, these runs included a range of particle sizes from 15 -- 60 cm ($\tau_{\rm s} = 0.25$ -- $1.0$) and a prescription for cooling by inelastic collisions was added.

To lessen the influence of initial conditions, the simulation was run in three phases.  First, the magneto-rotational turbulence developed for 10 orbits.  Second, particles were added with vertical gravity and two-way drag forces (similar to the back-reaction simulations in \S\ref{sec:KSclump}, but with higher resolution and a range of particle sizes).  After 10 orbits, vertical settling, turbulent clumping and streaming instabilities produced peak particle densities almost 100 times the gas density (slightly less than in \S\ref{sec:KSclump} due to the range of particle sizes), see JOMKHY Fig.\ 3.

Finally in the third phase, self-gravity and collisional cooling were turned on and the system ran for seven more orbital periods (see JOMKHY Fig.\ 2).  The densest clump of particles suddenly became a bound cluster of nearly a Ceres mass (equivalent to a 500 km planetesimal).  The cluster continued to accrete particles at $> (1/3) M_{\rm Ceres}$ per orbital period, for a ``final" (but still increasing) mass of $3.5 M_{\rm Ceres}$ after only seven orbits.

\subsection{Surprises}\label{sec:surprise}
There are many surprising features of the JOMKHY simulations which go against conventional wisdom. The Ceres mass scale of the bound clusters is striking, since it exceeds by many orders of magnitude the traditional estimate of a  ``km-sized planetesimal."  This mass need not go into a single planetesimal.  By analogy with Jeans collapse, one might expect further fragmentation due to collisional cooling during the collapse.  Only the initial stages of collapse have been modeled so far.

Also the Ceres mass scale is not a universal result.  The mass scale is introduced not by self-gravity acting on a smooth background as in GW, but rather by particle clumping in the (imposed and particle-generated) turbulent flow.  Thus the cluster mass depends on particle sizes and the nature of the imposed turbulence.

Indeed the supplementary information to JOMKHY \citep{supp07} considers the case of low ionization disks in \S12.1, where magetic fields are ignored, and only particle-driven turbulence exists.  The bound clusters in these simulations have masses and clump accretion rates lower by more than two orders of magnitude (see Tables 5 and 6 in the supplement).

This was perhaps the biggest surprise of the work, that imposed turbulence abets rather than hinders the gravitational collapse of solids.  The non-obvious explanation is that turbulent particle concentration has a "snowball effect." Particle clumping produced by magneto-rotational turbulence provides seeds where particle feedback produces denser clumps than the streaming instability (plus vertical shear) alone can trigger.

\section{Discussion}
The long-standing mystery of planetesimal formation suddenly appears much less daunting, if no less fascinating.  For decades, the obstacles seemed insurmountable: low sticking efficiencies, rapid radial migration and the inevitability of turbulent stirring.  Now it is clear that turbulent stirring does not necessarily prevent gravitational collapse of solid particles into planetesimals.  As discussed in \S\ref{sec:surprise}, more turbulence can promote gravitational collapse in some cases!

The breakthoughs in particle-gas dynamics are related to the role of particle feedback.  Once thought only to hinder particle settling by triggering Kelvin-Helmholz instabilities, we now realize that feedback triggers and augments particle clumping.  The link between feedback and clumping has  many pillars of support: the idealized streaming instability (YG, YJ, JY), Kelvin-Helmholz instabilities with uniform rotation \citep{jhk06}, and now 3D stratified models of Keplerian disk midplanes (JOMKHY and its supplement).

Much work and many uncertainties remain.  The large initial sizes assumed by JOMKHY might be unrealistic, either because coagulation stalls at smaller sizes, or if less violent gravitational collapse occurs first (see work on dissipative gravitational collapse by \citealp{war76,y05a}).  The fact that most primitive undifferentiated meteorites betray no structures larger than mm-sized chondrules suggest this concern may be valid.  However, particles which are small and tightly coupled to the gas disk are less effected by the dynamical clumping mechanisms discussed here, at least at the resolutions studied to date.  Perhaps more surprises are needed.

\acknowledgements %%% Text of acknowledgements runs on after this command.
Many thanks to the SOC and local hosts for a stimulating, and breathtaking, conference.  We acknowledge the contributions of our coauthors; Jeff Oishi, Mordecai Mac Low, Hubert Klahr, Thomas Henning, and Jeremy Goodman; to the work described here.

%%% THE BIBLIOGRAPHY
%%%
%%% CONSULT SECTION 3 OF "INSTRUCTIONS FOR AUTHORS" FOR HOW TO USE NATBIB.
%%% AUTHORS ARE ENCOURAGED TO USE EITHER THE "THEBIBLIOGRAPY" ENVIRONMENT
%%% BY UNCOMMENTING (DELETING THE "%" SYMBOL) THE COMMANDS BELOW, OR BY
%%% USING THE BIBTEX ENVIRONMENT. TO FIND OUT WHICH IS APPLICABLE TO YOUR
%%% CONTRIBUTION, CONSULT THE VOLUME EDITORS FOR YOUR PROCEEDINGS.
%%%

\end{document}